\documentclass[doublecol]{epl2}
\usepackage{gensymb}

\title{Intra-scales energy transfer during the evolution of turbulence in a trapped Bose-Einstein condensate}
\shorttitle{Intra-scales energy transfer during the evolution of turbulence in a trapped Bose-Einstein condensate} 

\author{Arnol Daniel Garc\'ia-Orozco\inst{1} \and Lucas Madeira\inst{1} \and
Luca Galantucci\inst{2} \and Carlo F. Barenghi\inst{2} \and Vanderlei S. Bagnato\inst{1,3}}

\institute{                    
  \inst{1} Instituto de F\'isica de S\~ao Carlos, Universidade de S\~ao Paulo, CP 369, S\~ao Carlos, S\~ao Paulo, Brazil\\
  \inst{2} Joint Quantum Centre Durham-Newcastle, School of Mathematics, Statistics and Physics, Newcastle  University, Newcastle upon Tyne, NE1 7RU,  UK\\
  \inst{3} Hagler Fellow, Department of Biomedical Engineering, Texas A\&M University, College Station, Texas 77843, USA
}


\abstract{
In turbulence phenomena, including the quantum turbulence in superfluids, an
energy flux flows from large to small length scales, composing a cascade of energy.
A universal characteristic of turbulent flows is the existence of a range of scales where the
energy flux is scale-invariant: this interval of scales is often referred to as inertial region.
This property is fundamental as, for instance, in turbulence of classical fluids it characterizes
the behavior of statistical features such as spectra and structure functions.
Here we show that also in decaying quantum turbulence generated in trapped Bose-Einstein condensates (BECs),
intervals of momentum space where the energy flux is constant can be identified. Indeed, we present
a procedure to measure the energy flux using both the energy spectrum and the continuity equation.
A range of scales where the flux is constant is then determined employing two distinct protocols and in the same
range, the momentum distribution measured is consistent with previous work.
The successful identification of a region with constant  flux in turbulent
BECs is a manifestation of the universal character of turbulence in these quantum systems. These
measurements pave the way for studies of energy conservation and dissipation in trapped atomic
superfluids, and also analogies with the related processes that take place in ordinary fluids.
}

\begin{document}

\maketitle

\section{Introduction}

Turbulence is a process that occurs in many types of fluids and in a wide range of length scales \cite{pope00}.
From the climatic effects involving large masses, down to capillaries, turbulence presents properties that are universal, regardless of the system under consideration. Richardson \cite{richardson22} was the first to propose that the phenomenology of three-dimensional turbulence is related to a forward cascade. An external source injects the kinetic energy at large length scales, thus feeding large eddies. The eddies interact, break up into smaller structures, and the process repeats itself down to small length scales, where energy is dissipated by viscous forces. It is convenient to work in momentum space when dealing with turbulence. Hence large (small) length scales correspond to small (large) scales in the reciprocal space. In order to avoid confusion with the word ``scale'', whenever we refer to the real space we write ``length scale'' explicitly, and ``momentum scale'' for the reciprocal space.

The existence of universal behavior in turbulence is considered a breakthrough in the understanding of the phenomenum. Among these universal characteristics, there is the so-called inertial region, usually represented in momentum space ($k$-space). In this region, energy must flow from small to large momentum scales of the system without loss of energy. The consequence of this universal feature was brilliantly elaborated in mathematical form by Kolmogorov \cite{zakharov12} and Obukhov \cite{monin54,foken06}.
Their work resulted in one of the best-known laws in the field of turbulence: the kinetic energy spectrum has a power-law behavior,
\begin{equation}
E(k)\propto k^{-\delta},
\end{equation}
where $\delta$ a fractional or integer number, depending on the type of turbulence \cite{barenghi14}. This characteristic, which is a universal behavior for turbulence phenomena involving classical fluids, has also been observed in superfluids, when in the quantum turbulence regime \cite{kobayashi05,baggaley12}.
Initially, this behavior was investigated in liquid He, both from theoretical and experimental point of views.
More recently, this was also investigated in atomic superfluids that arise during Bose-Einstein condensation of trapped atoms \cite{thompson13,navon16,vivanco17,tsatos16,madeira20}.

In the inertial interval, the energy flows between momentum scales in a non-dissipating regime,
therefore it yields a constant energy flux.
The presence of turbulence forces the energy injected in small momentum scales to flow in a constant flow regime to large momentum scales, where it is dissipated.
This description seems entirely appropriate for a wide variety of turbulent phenomena.

The presence of a constant energy flow regime seems to be as important as the power-law itself, or perhaps even more, in terms of identifying the turbulence phenomenon as it is one of the expected characteristics for such non-equilibrium states in fluids.
A region with constant flux suggests the existence of a cascade, while
the power-law exponent contains details about it
(waves, eddies, a mixture of both).
However, because it is easier to measure the energy spectrum and the presence of the power-law, this has been the main way to identify the presence of turbulence. Obtaining the energy spectrum and identifying simultaneously the energy flux in different momentum ranges may bring great physical insight into the experiments.

These concepts have been used in investigations of related systems.
For example, the energy flux was used to present evidence of an inverse energy transfer, induced by vortex reconnections, in numerical simulations of superfluid helium \cite{baggaley14}.
From the experimental point of view, 
Navon \textit{et al.} \cite{navon19} used particle and energy fluxes to investigate a turbulent quantum gas. The main differences with respect to our present study is that they employed a cylindrical box trap (leading to a homogeneous Bose gas), and they were interested in steady-state dynamics, while we focus on the decaying regime of turbulence.

In this work, experiments with an $^{87}$Rb Bose-Einstein condensate were used to produce the quantum turbulence state by introducing excitations as in a previous works \cite{henn09,henn09_2,henn10,seman11,shiozaki11}.
Since we are interested in the decaying turbulence regime, the system is perturbed and we study the time evolution after the external excitations are turned off. 
Once the desired state is produced, the momentum spectrum is obtained by an optical absorption image after a free time of flight (TOF), that corresponds to a projection of the cloud density on a plane.
Then the energy spectrum is obtained, considering a kinetically dominated regime. The final energy spectrum is analyzed to identify regions where flux remains constant
or regions where the total energy does not change.

This work is organized as follows.
We start by introducing some concepts related to the energy flux
that will be applied when analyzing the measurements.
Then, a brief description of the experimental procedure is offered, followed by the results. Finally, we discuss our findings and their implications.

\section{Energy flux and cascade in quantum turbulence}

Let us consider a system, where the excitations are introduced in a scale with characteristic wave number $k_D=2\pi/D$, where $D$ is the typical system size. Due to the mechanisms involved in superfluids, interactions occur between the excitations, whether they are vortices or waves. The result of these interactions is that smaller structures are formed, and energy begins to flow from the large to the small length scales in a process called energy cascade \cite{zakharov12}.
Initially, at large length scales, the system is forced by injecting energy in the form of large structures composed of vortices or waves.
A process of forming excitations on smaller length scales occurs, entering the inertial region where well-known cascades such as Kolmogorov's, Kelvin's ones\cite{vinen01,barenghi14_2}, or others can occur. In this scale range, energy only migrates from one scale to another, without dissipating, until it finds a smaller length scale where dissipation begins to occur.
This is a way of the system to escape the original excited state.
Energy transfer is a typically out-of-equilibrium process, while at a micro-scale, the system is in a near-equilibrium regime. Thus, the turbulence process is a phenomenon of multiple scales, which evolves in time seeking an equilibrium condition.

Denoting by $\boldsymbol{\Pi}_E(\textbf{k},t)$ the energy flow (energy per unit area and per unit time in $\textbf{k}$-space), $I_E(\textbf{k},t)$ the rate of external energy injection producing excitations, $D_E(\textbf{k},t)$ the energy dissipation rate for a given wave number $\textbf{k}$ at time instant $t$, we can write the continuity equation relating those quantities,
\begin{equation}
\frac{d\rho(\textbf{k},t)}{dt}=  I_E (\textbf{k},t)- D_E (\textbf{k},t)- \nabla_\textbf{k} \cdot \boldsymbol{\Pi}_E(\textbf{k},t),
\end{equation}
where $\rho(\textbf{k},t)$ is the energy density.
We are interested in the regime of decaying turbulence, after the energy input has stopped, hence $I_E (\textbf{k},t)=0$.
Also, the dissipation occurs at $k$-values larger than the ones corresponding to the inertial range.
If we are dealing with a momentum region where injection and dissipation are not present, the expression is simplified,
\begin{equation}
\label{eq:vector_cont}
\frac{d\rho(\textbf{k},t)}{dt}+\nabla_\textbf{k} \cdot \boldsymbol{\Pi}_E(\textbf{k},t)=0.
\end{equation}
If we assume an isotropic energy distribution, Equation (\ref{eq:vector_cont}) can be written as a function of the modulus of $\textbf{k}$, $|\textbf{k}| = k$, resulting in
\begin{equation}
4\pi k^2 \frac{d\rho(\textbf{k},t)}{dt}=-\frac{\partial \Phi_E(\textbf{k},t)}{\partial k},
\end{equation}
where we introduced the scalar flux $\Phi_E(\textbf{k},t)$ corresponding to an
isotropic flux vector field $\boldsymbol{\Pi}_E(\textbf{k},t)$.
By definition, $4\pi k^2 \rho(k,t)=E(k,t)$, where $E(k,t)$ is the energy spectrum, such that $\int E(k)dk$ equals the total energy of the system.
Hence, the  energy flux in momentum space can be written as
\begin{equation}
\label{eq:flux}
\Phi_E(k,t)=-\int_{k_D}^k \frac{dE(k',t)}{dt} dk'= -\frac{d}{dt} \int_{k_D}^k E(k',t)dk'.
\end{equation}
Equation (\ref{eq:flux}) indicates that observing the time variation of the energy
spectrum in a specific interval, from its smallest value $k_D$ to a given value of momentum $k$, one can obtain the energy flux at $k$. This feature is the cornerstone of this work, and this expression for the energy flux has been previously used by Baggaley and co-authors in the context of quantum turbulence \cite{baggaley14}.

\section{Obtaining a turbulent cloud in a trapped superfluid}

The experiment begins with the production of a Bose-Einstein condensate, containing about $4\times 10^5$ $^{87}$Rb atoms in the hyperfine state
$|F, m_F \rangle=| 2,2 \rangle$, confined in a Quadrupole-Ioffe magnetic trap. At the end of the experimental procedure, the condensate fraction is approximately 80\%. Details of the experimental condensate production and other technical remarks can be found in
Refs.~\cite{henn10,seman11,shiozaki11}.

After the condensate is produced, while it is still in the trap, an oscillating magnetic field is applied. The field is produced by a pair of anti-Helmholtz coils placed with their axis tilted by a small angle, of approximately 5\degree, with respect to the axis of the trap.
Since the coils are not aligned with the condensate axis, the oscillations generate deformations, displacements, and rotations in the cloud.
The amplitude of the disturbances introduced by the excitation coils can be varied, as well as the time and frequency of excitation. In this work, we keep the excitation frequency fixed at $\Omega/(2\pi)=$ 190 Hz. The excitation amplitude, measured by the gradient of the input field, and quantified by the voltage in a standard resistor, is varied until an amplitude is reached, where changes in the cloud are clear enough to guarantee the establishment of a non-equilibrium state with turbulent characteristics.
The excitation time may also vary because there is a compromise between the excitation application time and the amplitude to generate the turbulent state. For example, larger amplitudes need less time to reach similar conditions.
The range of amplitudes to obtain turbulence was the topic of investigation
in previous works \cite{henn10,seman11,shiozaki11}.

In this paper, we employed an excitation time $\tau_{\rm exc}$ of 35 ms.
Our temporal resolution is of approximately 0.1 ms.
After the excitation is done, the coils that generate the disturbance field are turned off, and the system is left on hold for a time
$\tau_{\rm hold}$.
This time may vary from 20 to 70 ms. During this interval, there is a temporal evolution of the state produced during the excitation period.
During the hold period, the momentum distribution evolves in time. The variation of $\tau_{\rm hold}$ allows the determination of the temporal variation of the energy spectrum and, consequently, the possible existence of momentum intervals where a constant flow is present.

\begin{figure}[!htb]
\begin{center}
\includegraphics[width=0.9\linewidth]{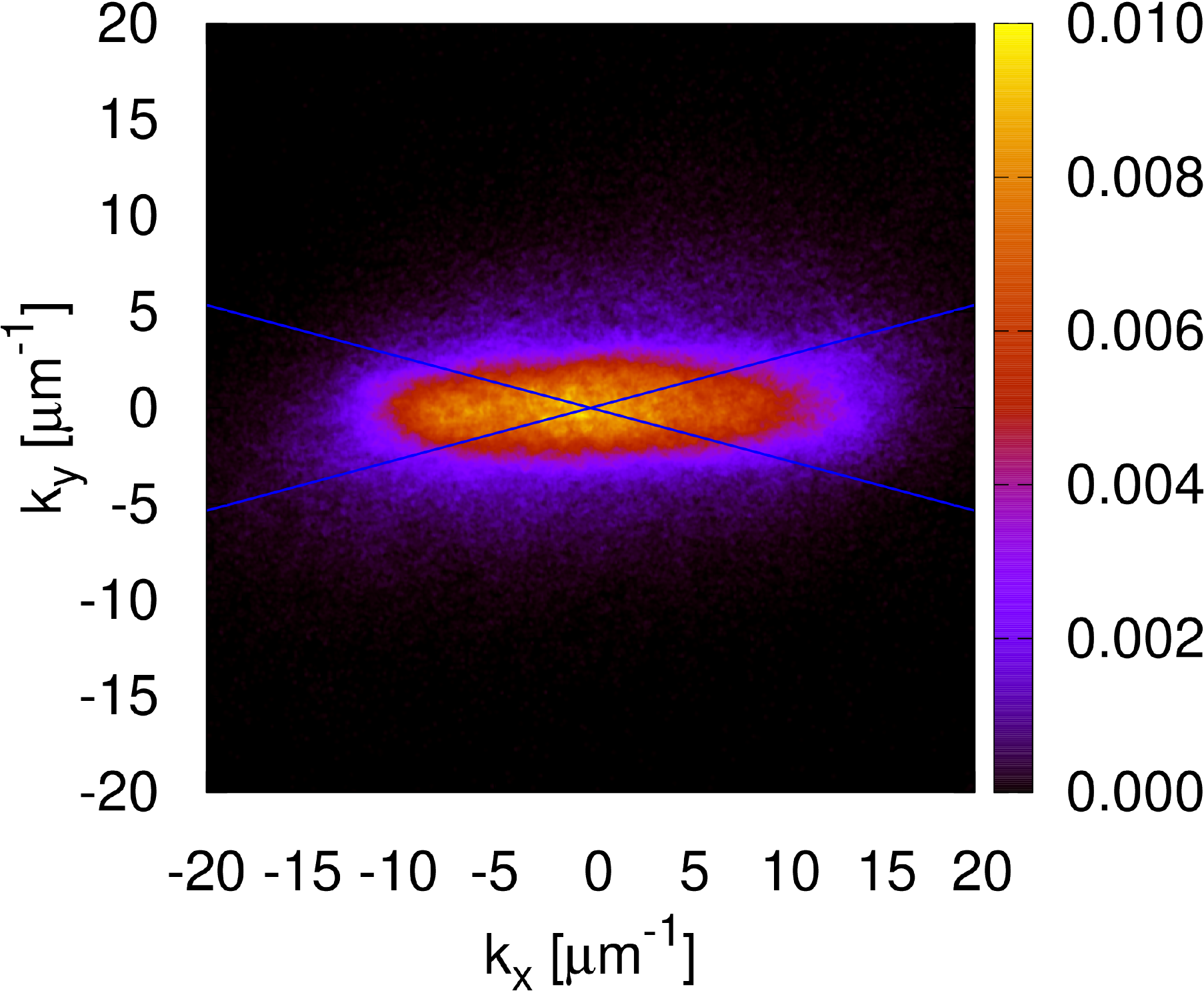}
\caption{In-plane momentum distribution of the turbulent cloud with 
$\tau_{\rm hold}=$ 46 ms. The normalization is such that the integration over the plane yields one. The two blue lines correspond to $k_y=\pm (\tan 15^\circ)k_x$, i.e., two 30$^\circ$ apertures centered around the major axis of the expanded cloud.
}
\label{fig:cloud}
\end{center}
\end{figure}

After a particular state is produced, and the hold time has passed, the atoms are released from the trap and perform a free expansion during a time $\tau_{\rm TOF}$.
As previously investigated \cite{caracanhas13}, the turbulent state is kinetically dominated, hence the interaction energy 
cannot be responsible for the majority of the effects in a turbulent cloud.
The momentum distribution for a non-interacting BEC can be obtained by analyzing the density distribution of the cloud after a finite TOF.
Following the release of the trap, the atoms expand ballistically \cite{dalfovo99}. After a time $\tau_{\rm TOF}$, the distance that an atom has traveled from the center of the trap is given by $r=\hbar\tau_{\rm TOF}k/m$, where $\hbar$ is Planck's constant and $m$ is the atomic mass.
Hence, the spatial distribution of atoms in free expansion can be used to map the momentum distribution, $n(r)=n(\hbar\tau_{\rm TOF}k/m)$, and the
momenta of the atoms are connected with the position in the expanded cloud.
The validity of the application of the time of flight technique to obtain the momentum distribution has been widely discussed in several publications \cite{thompson13,navon16}.
The procedure we adopted is to perform an optical absorption
image of the cloud after the time of flight. This produces an image on the plane $(k_x, k_y)$, as illustrated in Fig.~\ref{fig:cloud}.

Equation~(\ref{eq:flux}) was derived assuming an isotropic momentum distribution, because that is the simplest model one can assume, which is clearly not the case for our experiment.
In order to investigate the impact of the anisotropy, throughout this work we will consider the application of our model to two cases:
(i) the entire cloud and (ii) only the regions close to the major axis of the expanded cloud, i.e., the two regions bounded by $k_y=\pm (\tan 15^\circ)k_x$, which we depict as blue lines in Fig.~\ref{fig:cloud}.
The momentum distribution is obtained, in both cases, by averaging the number of atoms within the interval between $k=\sqrt{k_x^2+k_y^2}$ and $k+\delta k$, with $\delta k\approx$ 0.05 $\mu$m$^{-1}$.
The integration of the in-plane density, for a given time $t$, yields the number of particles in that region,
\begin{equation}
\int d\Omega \ dk \ k^2 n(k,t)=N(t),
\end{equation}
where the integration over the solid angle yields $4\pi$ for the whole cloud, or $2\pi/3$ for the regions close to the major axis that we defined.
Since the total number of atoms is approximately constant comparing different
time intervals, variations of $n(k,t)$ are due to effects arising from the time evolution of the turbulent state alone.

\begin{figure}[!htb]
\begin{center}
\includegraphics[width=0.7\linewidth,angle=-90]{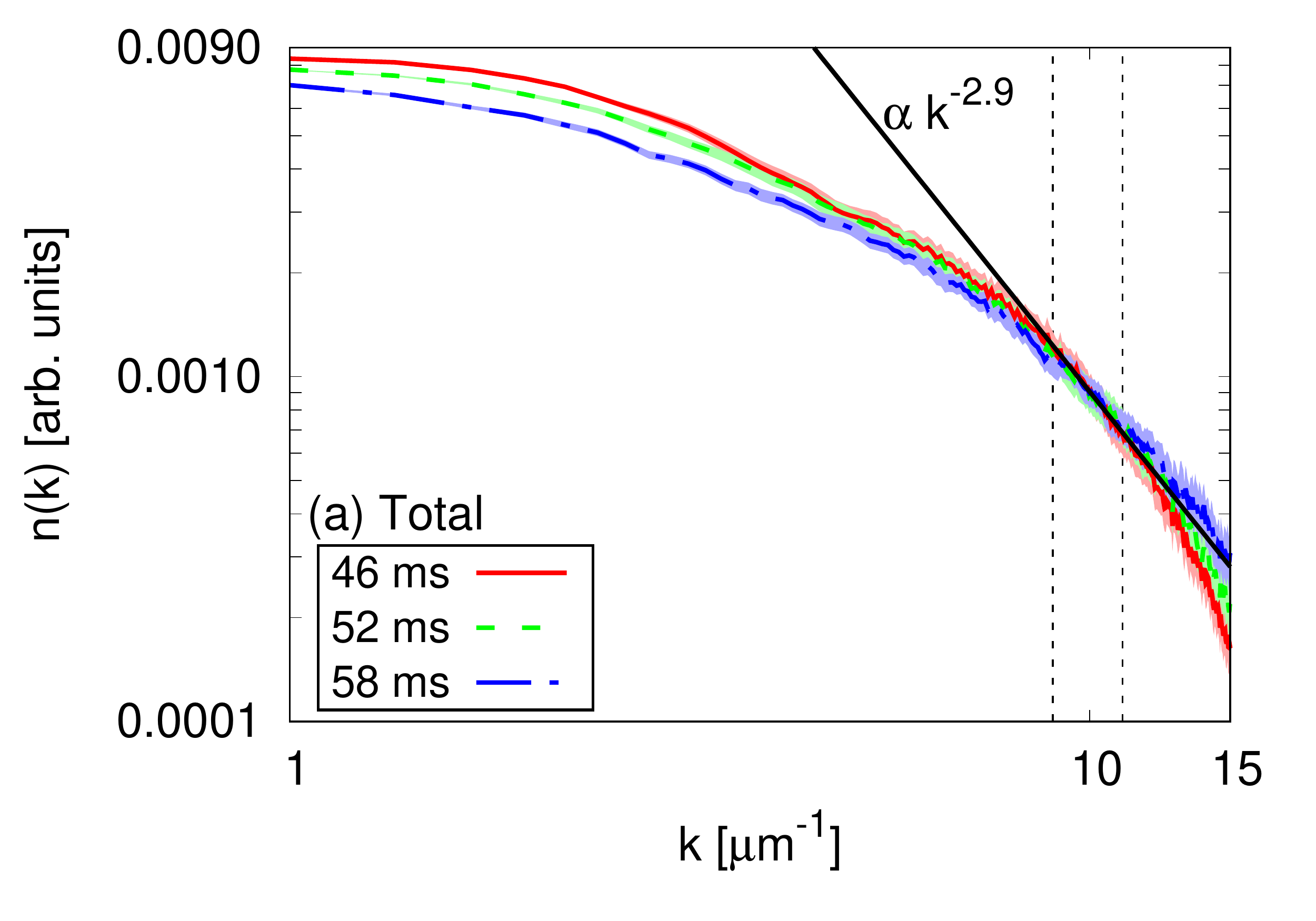}
\includegraphics[width=0.7\linewidth,angle=-90]{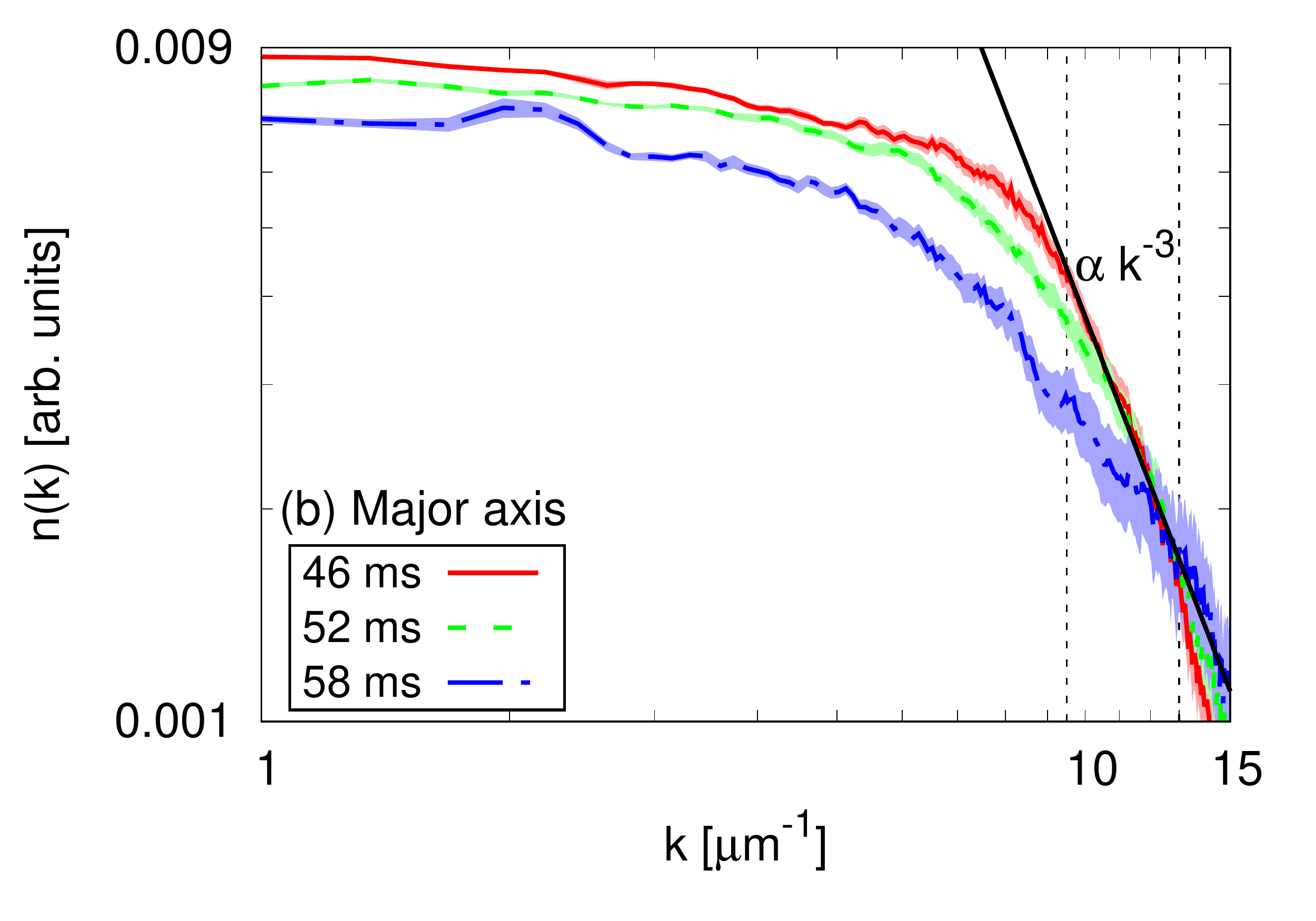}
\caption{Momentum distribution obtained for different holding times,
$\tau_{\rm hold}=$ 46, 52, and 58 ms denoted by the red solid curve, green dashed curve, and blue dash-dotted curve, respectively.
The upper panel corresponds to an average of the whole cloud, whereas we consider only the region close to the major axis, $k_y=\pm (\tan 15^\circ)k_x$, in the lower panel.
The uncertainties are represented by the shaded areas around the curves.
As time progresses, the distribution modifies itself, decreasing for lower momenta and increasing for higher ones.
The solid black lines correspond to a power-law $n(k)\propto k^\delta$ with $\delta=-2.9$ and -3.0 for the upper and lower panels, respectively.
The dashed lines denote the range where the power-law is observed for $\tau_{\rm hold}=$ 46 ms. For the other values of $\tau_{\rm hold}$ the same exponent is obtained for a slightly higher $k$-range.}
\label{fig:nk}
\end{center}
\end{figure}

Figure~\ref{fig:nk} shows typical momentum distributions obtained for the turbulent cloud, produced under the described conditions.
The momentum range available in the experiment corresponds to 30 $\mu$m$^{-1}$, but, for simplicity, we only plot the first half of the range.
As the holding time increases, the distribution shifts towards higher momenta.
We present the distributions both for an angular averaging procedure which takes into account the whole cloud, Fig.~\ref{fig:nk}(a), and only the regions close to the major axis, Fig.~\ref{fig:nk}(b).

For $\tau_{\rm hold}=$ 46 ms
we performed a fit in the 9 $\le k \le$ 11 $\mu$m$^{-1}$ range (denoted by dashed lines in Fig.~\ref{fig:nk}(a)), considering the whole cloud, to a power-law which yielded the exponent -2.9(1), in agreement with previous experiments \cite{thompson13}. The same exponent is observed for $\tau_{\rm hold}=$ 52 and 58 ms, however, for slightly higher momenta, 10 $\le k \le$ 12 $\mu$m$^{-1}$.
We applied the same procedure to the results corresponding to the averaging near the vicinity of the major axis of the cloud. We found an exponent of -3.0(1) for $\tau_{\rm hold}=$ 46 ms in a broader range than before, 9.5 $\le k \le$ 13 $\mu$m$^{-1}$, indicated by dashed lines in Fig.~\ref{fig:nk}(b). The same exponent is also observed for $\tau_{\rm hold}=$ 52 and 58 ms, again, for a higher momentum range.

Some considerations regarding the two angular averaging procedures are in order. Averaging the whole cloud introduces contributions at low momenta, if compared to considering only the regions in the vicinity of the major axis, as can be inferred from Fig.~\ref{fig:cloud}.
This effect has two consequences in the power-law behavior of $n(k)$.
Firstly, the $k$-range where the power-law scaling is observed occurs at slightly lower momenta, 9 $\le k \le$ 11 $\mu$m$^{-1}$, for the entire cloud with respect to the region close to the major axis, 9.5 $\le k \le$ 13 $\mu$m$^{-1}$, for a given $\tau_{\rm hold}$. Secondly, the region is slightly broader for the major axis case.
Remarkably, the exponents are very close in the two cases, which allows us to compare the two cases in order to better understand the role of anisotropy in this system.

\section{The energy spectrum and the analysis of the energy flux}

\begin{figure}[!htb]
\begin{center}
\includegraphics[width=0.6\linewidth,angle=-90]{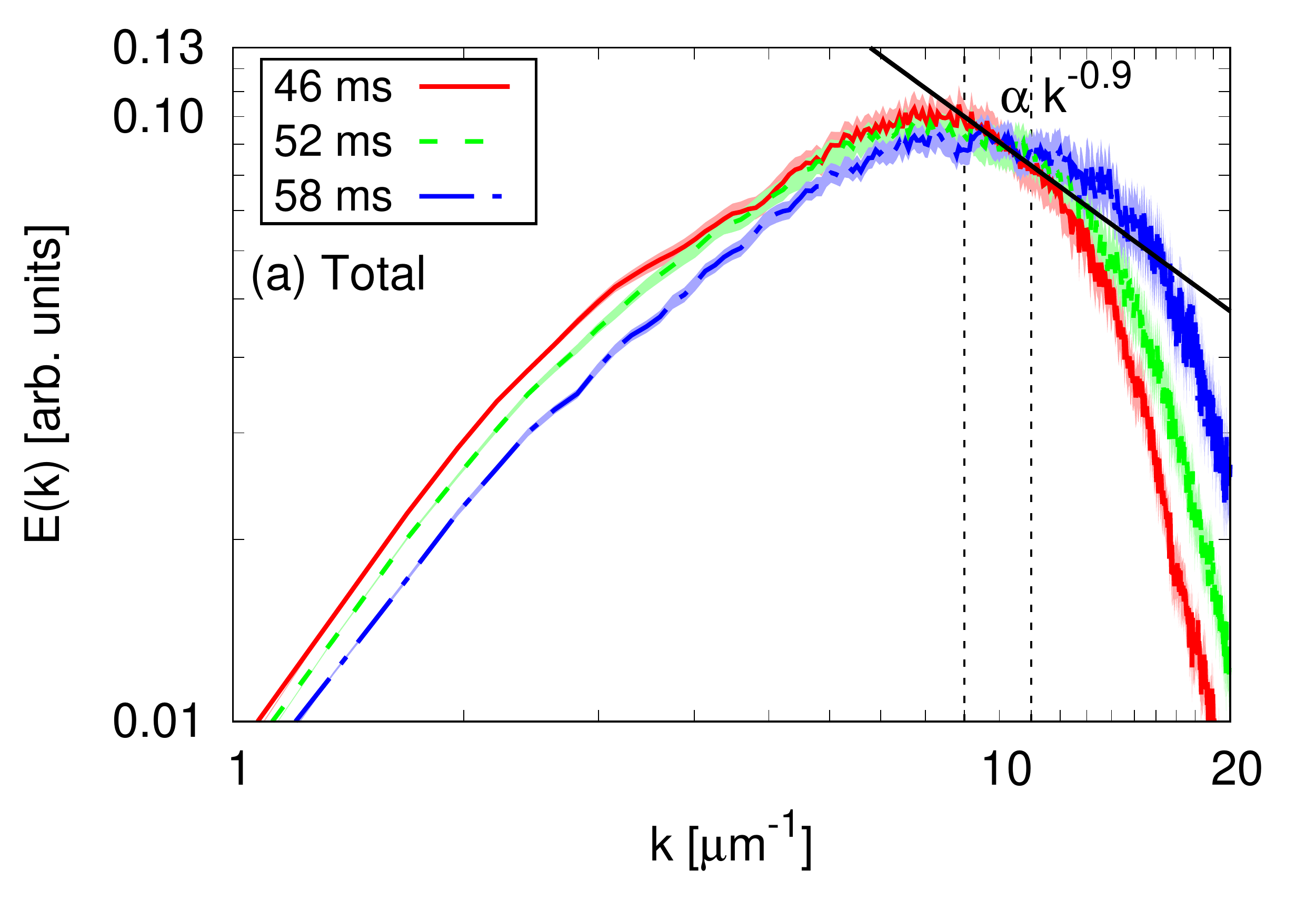}
\includegraphics[width=0.6\linewidth,angle=-90]{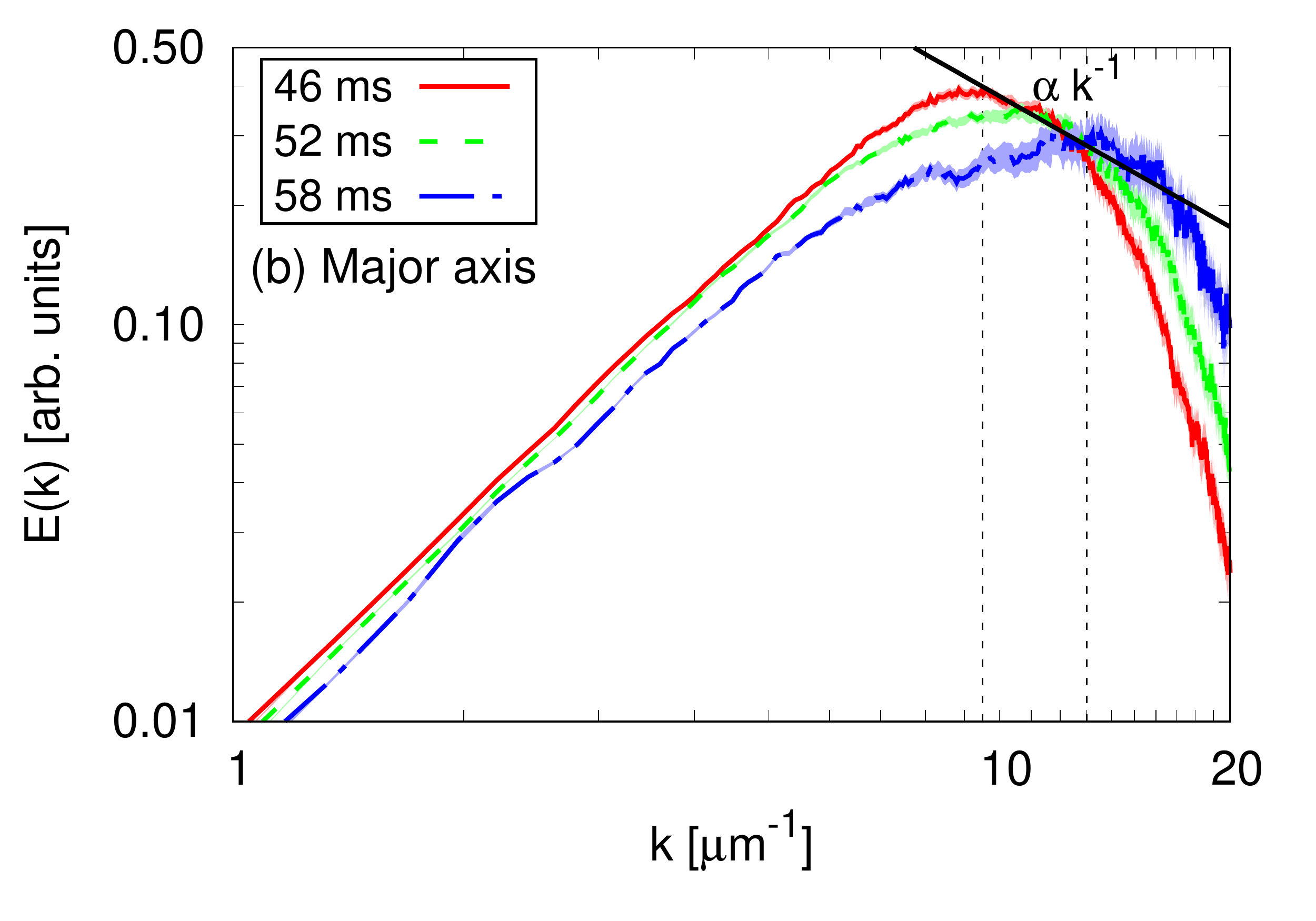}
\caption{Energy spectrum $E(k)$ for the turbulent cloud for different holding times, $\tau_{\rm hold}=$ 46, 52, and 58 ms.
The curve styles are the same as the ones employed in Fig.~\ref{fig:nk}.
The upper panel corresponds to an average of the whole cloud, whereas we consider only the region close to the major axis in the lower panel.
The evolution
of the spectra is toward migration of energy from lower to higher momenta, as expected.
The solid black lines correspond to a power-law $E(k)\propto k^\delta$ with $\delta=-0.9$ and -1.0 for the upper and lower panels, respectively.
The dashed lines denote the range where the power-law is observed for $\tau_{\rm hold}=$ 46 ms.
}
\label{fig:ek}
\end{center}
\end{figure}

From the momentum distribution, the energy spectrum can be obtained by multiplying $n(k)$ by the energy of each component, $\hbar^2 k^2/(2m)$.
Hence, the power-law behavior of the energy spectrum is observed in the same regions as the ones for the momentum distribution, with an exponent that differs by 2 (due to the $k^2$ in the definition).
The energy spectra obtained for the 46 ms, 52 ms, and 58 ms holding times are shown in Fig.~\ref{fig:ek}, both for the entire cloud (a), and the region close to its major axis (b). Note that with increasing holding time, the energy migrates from the small to large $k$ values, as expected.

\begin{figure}[!htb]
\begin{center}
\includegraphics[width=0.6\linewidth,angle=-90]{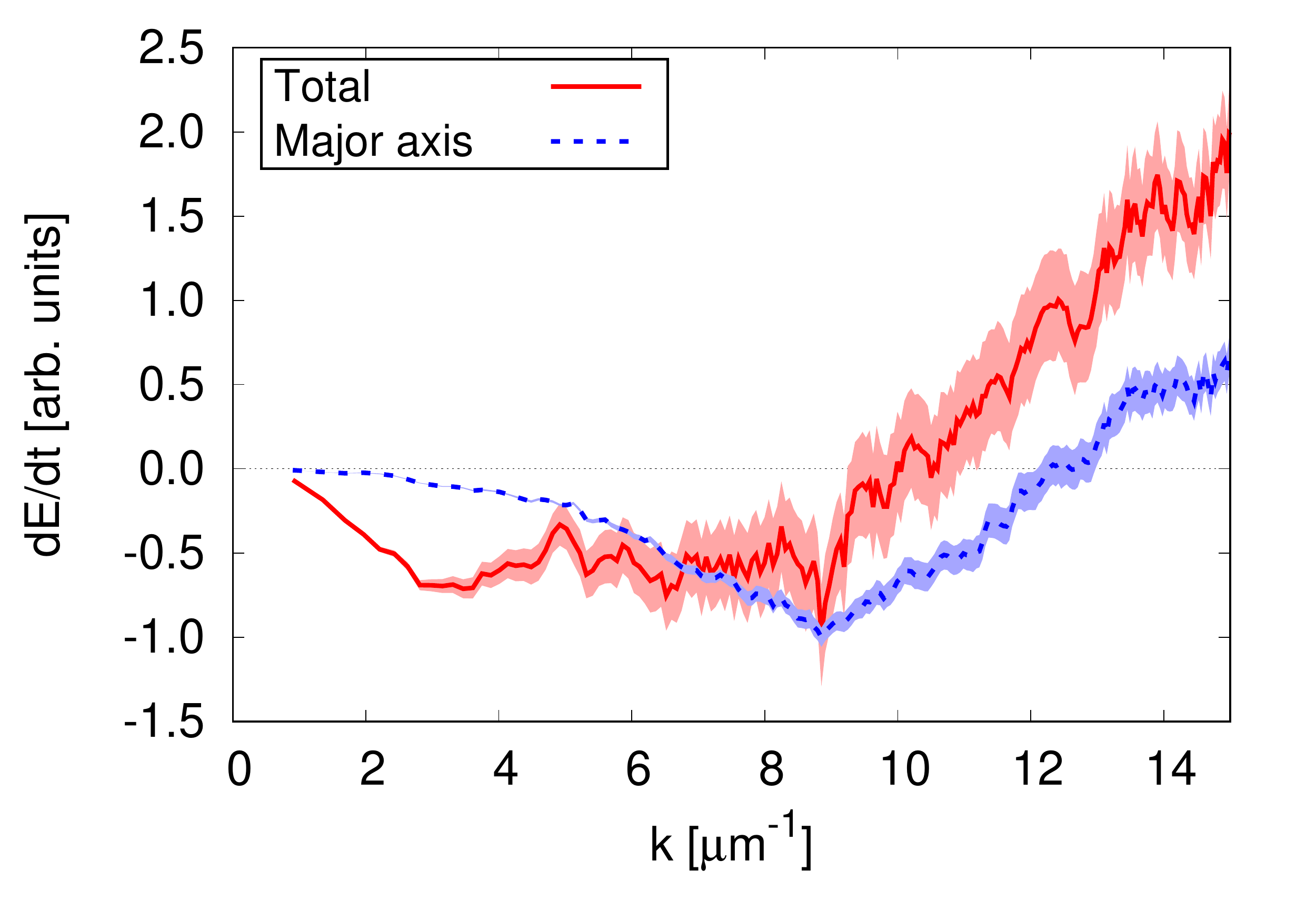}
\includegraphics[width=0.6\linewidth,angle=-90]{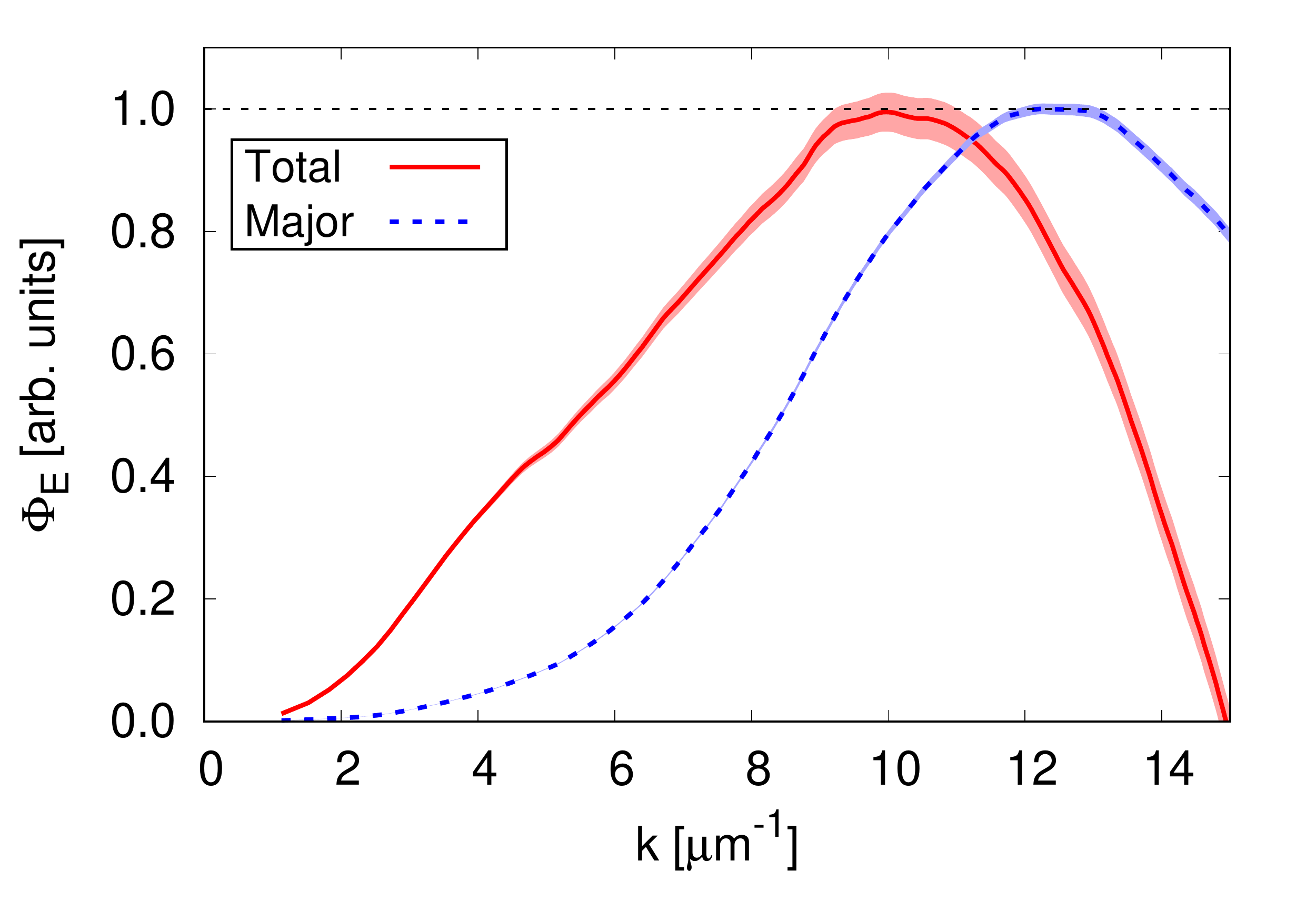}
\caption{Time derivative of the energy spectrum (upper panel) and corresponding energy flux (lower panel).
The solid red curves denote an average of the whole cloud, whereas the dashed blue curves correspond only to the regions close to the major axis.
The shaded regions correspond to the uncertainty.
The fact that the energy derivative with respect to time is negative for small values of $k$ and positive for large ones indicates that energy
is flowing from small to large momenta.
Notice an approximately constant flux from $\approx$ 9 to 11 $\mu$m$^{-1}$ if we consider the entire cloud, and $\approx$ 11.5 to 13.5 $\mu$m$^{-1}$ for the regions close to the major axis of the expanded cloud.
}
\label{fig:flux}
\end{center}
\end{figure}

With the spectra at three different times, one can determine the flux
$\Phi_E(k, t)$ previously defined in Eq.~(\ref{eq:flux}).
We adopted the flow evaluation at 52 ms, using other holding times to determine the time derivative of the energy spectrum, and its associated uncertainty. We performed the
numerical differentiation using a three-point equally-spaced abscissas formula
\cite{abramowitz70}. The time derivative, and its associated uncertainty,
is displayed in the left panel of Fig.~\ref{fig:flux}.
The calculation is performed for each value of $k$, taking $k_D=2\pi/D$ with $D \approx 30$ $\mu$m. The resulting flux is shown in the right panel of Fig.~\ref{fig:flux}.

Again, it is possible to see the differences between averaging the entire cloud, or just the regions close to its major axis.
Excluding the low momenta contributions of averaging the entire cloud
reduces the variance of the quantities and also shifts the constant flux region to higher momenta.
The fact that the flux is positive implies that the energy is flowing from the smallest to the largest $k$ values characterizing the direct cascade of energy, see Fig.~\ref{fig:flux}.
It is observed that the flux, $\Phi_E(k, t = 52$ ms), grows from small values of $k$, increasing until reaching a particular value, where it begins to level out, then acquires an approximately constant value, after which it begins to decrease again.
This region of constant flux corresponds to
$\approx$ 9 to 11 $\mu$m$^{-1}$ if we consider the whole cloud, and $\approx$ 11.5 to 13.5 $\mu$m$^{-1}$ for the regions close to the major axis.
For the region with constant flux, the migration of energy between the momentum scales (although we do not have a broad range) occurs without energy being added or subtracted.
A previous investigation in similar conditions\cite{thompson13} found that
$n(k)$ displays a power-law dependence in the range between 5 and 20 $\mu$m$^{-1}$, customarily associated with the inertial range, characterizing the possible cascade.
The $k$-range we found in this work is within this previous estimate.

\begin{figure}[!htb]
\begin{center}
\includegraphics[width=0.6\linewidth,angle=-90]{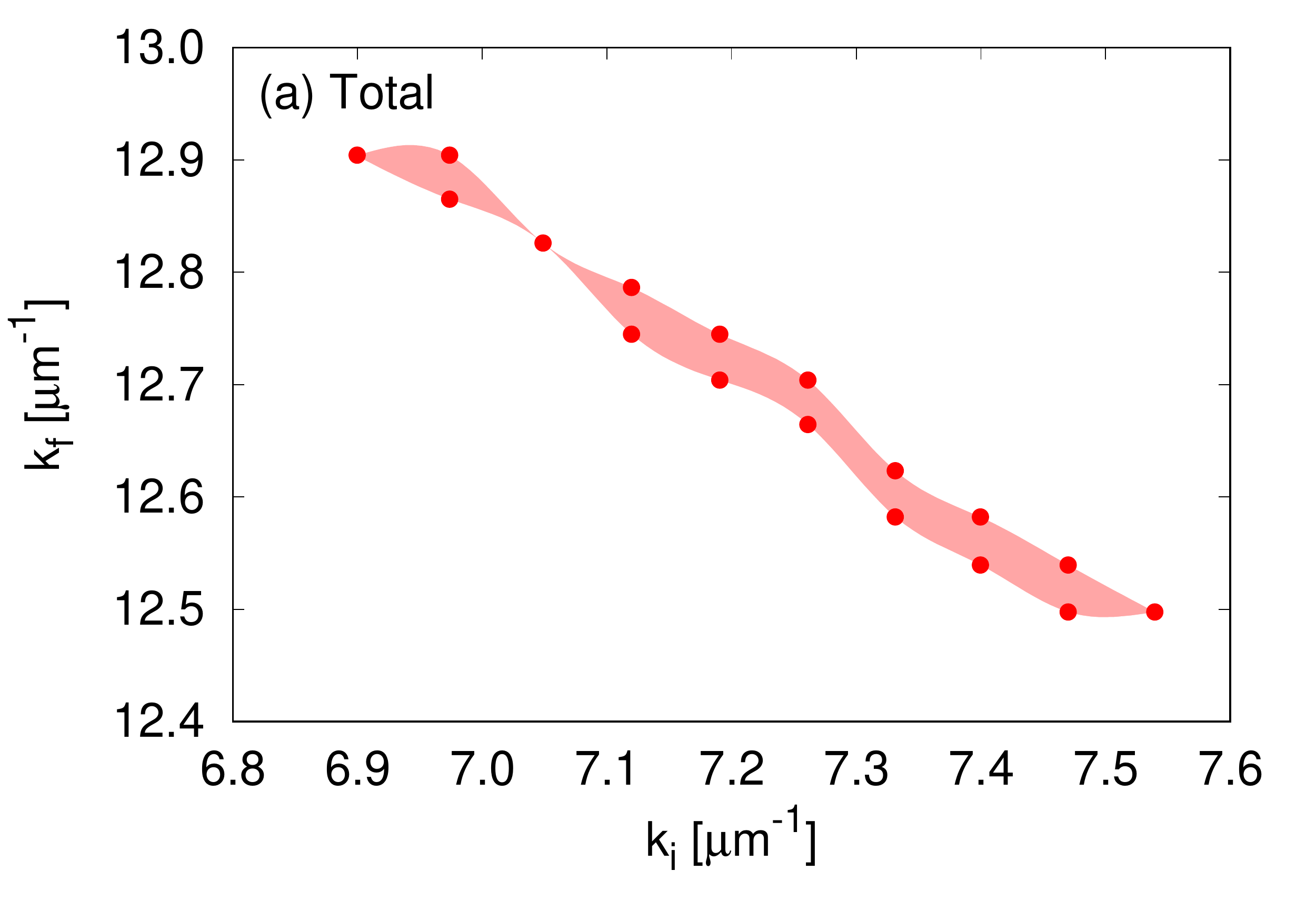}
\includegraphics[width=0.6\linewidth,angle=-90]{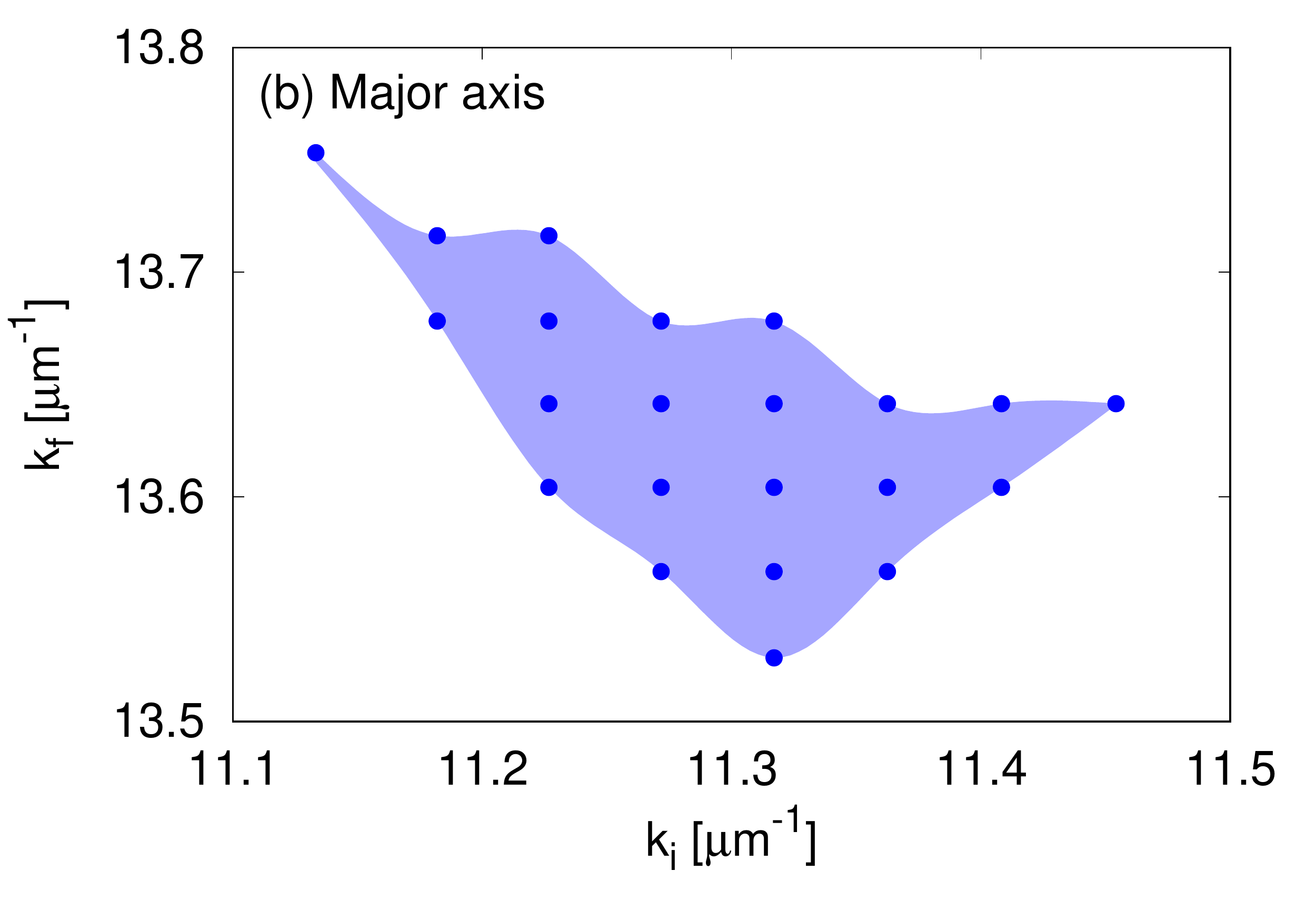}
\caption{Intervals $[k_i,k_f]$, as defined in Eq.~(\ref{eq:Q}), corresponding to $Q(k_i, k_f,46$ ms)$\approx Q(k_i, k_f,52$ ms)$\approx Q(k_i, k_f,58$ ms).
The shaded regions are plotted just to guide the eye.
The upper panel corresponds to an average of the entire cloud (a), and the lower panel to the regions close to the major axis (b), $k_y=\pm (\tan 15^\circ)k_x$.
This suggests that there is no energy accumulation in the $k$-range $7.2(4)\le k \le 12.7(2)$ $\mu$m$^{-1}$ in the first case, and $11.3(2)\le k \le 13.6(2)$ $\mu$m$^{-1}$ in the second case.}
\label{fig:q}
\end{center}
\end{figure}

The existence of a region characterized by a constant energy flux $\Phi_E (k,t)$ is confirmed by the fact that there is an interval in $k$-space where the total energy (the integral of the spectrum) is preserved, i.e. without variation with time evolution. This happens because no energy is accumulated or dissipated in this region, as a direct consequence of constant energy flux. In order to identify this region of preserved total energy, we look for intervals in $k$ space where the integral of the spectrum is constant for different times.
We define two limits, $k_i$ and $k_f$, where there is no temporal dependence for the integral of the energy spectrum. We start by defining the quantity
\begin{equation}
\label{eq:Q}
Q(k_i,k_f,t)=\int_{k_i}^{k_f} E(k,t)dk.
\end{equation}
The idea is to look for the interval $[k_i, k_f]$ that has independence over time.
From the curves $E(k,t)$, we perform a numerical calculation, determining the interval in which $Q(k_i, k_f,52$ ms) does not differ from $Q(k_i, k_f,46$ ms) or $Q(k_i, k_f,58$ ms) by more than an arbitrary and small value. The result is shown graphically in Fig.~\ref{fig:q} for the two angular averaging procedures. 

It is observed that for the given spectra, the interval where there is a minimal temporal variation of $Q(k_i,k_f,t)$ is between 7.2(4) to 12.7(2) $\mu$m$^{-1}$ if we average the entire cloud, and $11.3(2)\le k \le 13.6(2)$ $\mu$m$^{-1}$ if we only consider the regions near the major axis. 
Again, this coincides with the previous constant flux interval and also with the interval where the existence of a power-law in the momentum distribution was determined, characterizing the energy cascade.
Moreover, the variance of the values of $k_i$ and $k_f$ is smaller in the case where we consider only the vicinity of the major axis.

\section{Conclusions and final remarks}

In conclusion, we have presented a procedure to identify the inertial range in decaying quantum turbulence that is much more clear than looking for a power-law in the small range of momentum distribution which is available in experiments.
By obtaining the energy spectrum of a turbulent cloud of condensed atoms, it was possible to identify that there are regions in the reciprocal space where the energy flux between the momentum classes is independent of $k$ in the decaying turbulence regime. This characterizes an inertial regime where energy flows between momentum scales without dissipation.

The identification of this region from the momentum distribution alone would be very difficult, see Fig.~\ref{fig:nk}.
However, the identification is much more clear if we employ the
energy flux illustrated in Fig.~\ref{fig:flux}(b),
with the results confirmed by the identification of the range $[k_i,k_f]$ defined by Eq.~(\ref{eq:Q}), displayed on Fig.~\ref{fig:q}.
These two methods also provide intervals that are close together,
$\approx$ 9.5 to 13 $\mu$m$^{-1}$ and
$11.3(2)\le k \le 13.6(2)$ $\mu$m$^{-1}$ in the case where we consider only the vicinity of the major axis of the expanded cloud; and
$\approx$ 9 to 11 $\mu$m$^{-1}$ and
$7.2(4)\le k \le 12.7(2)$ $\mu$m$^{-1}$ for the entire cloud.
These intervals agree with previous measurements
of the power-law behavior of the momentum distribution \cite{thompson13}.

Classical turbulence is possible in anisotropic systems so we believe it is also important to study it in quantum systems. For this reason we computed all the quantities in this work using two different procedures: an angular average of $n(k)$ considering the entire expanded cloud, and only the regions close to its major axis.
We found that the low-momenta contributions outside the region of the major axis pull the inertial range toward slightly lower values.
What is interesting is that the procedure we adopted is consistent independently of the momentum distribution we use as input, what changes is the inertial range that we find.

It is worth mentioning that our system is different from those of previous works, for example Ref.~\cite{navon19}, where the interest was on pre- and steady-state dynamics, and not the decaying turbulence.
Although the trapping potential employed in this work is different from the one of Ref.~\cite{navon19}, which is reflected in different power-law exponents, in both situations it is possible to observe scale-invariant fluxes. We believe this evidence shows the importance of energy and particle fluxes in the study of quantum turbulence.

The range we found is relatively short, with less than one order of magnitude. However, this is characteristic of the system used, where the range of available scales is relatively short with less than one-decade order available. The range found, is compatible with previous measurements where in the same range is, for the energy spectrum or the momentum distribution, a region with a power-law of dependence with $k$.

Visualization techniques are well-developed in liquid helium systems \cite{bewley06},
and the same level of detail has not been achieved in trapped condensates yet, thus much of the progress has been
done relying on numerical simulations.
In this sense, simulations matching our experimental setup would help to test our assumptions and to validate our findings.

Our results complement previous findings by contributing more facts in characterizing turbulent state in excited Bose-Einstein condensates. Despite this fact, it should be noted that our sample is three-dimensional, and the ideal situation would be an exploration of the momentum and energy spectrum in 3D, without the need to perform the projection on the plane. The in-plane projection of the expanded cloud mixes high and low momentum components, limiting the observations.
If we were to perform another projection of the momentum, now obtaining an in-axis momentum, we would even observe a further reduction of the interval of constant flux, the mixture being even more severe.
An estimate of the expected range for the 3D cloud, if we could measure it, would result in a range of $k_i$ to $k_f$ $\approx$ 7 $\mu$m$^{-1}$, a broader range than that observed in this paper. This is associated with the necessity to project the cloud on the plane.  At the moment, we have to manage the limitations of the technique and work to overcome those limitations. In any case, the results are interesting and indicative of the existence of the energy cascade with the presence of a region in momentum space for a constant flux of energy in the system.
The results presented here open up new possibilities to investigate the
intra-scales aspects of the energy flux in quantum turbulence, and new experiments are on the way.

\acknowledgments
This work was supported by
the S\~ao Paulo Research Foundation (FAPESP)
under the grants 2013/07276-1, 2014/50857-8, and 2018/09191-7, and by the
National Council for Scientific and Technological Development (CNPq)
under the grant 465360/2014-9.
L.G. and C.F.B. acknowledge the support of the Engineering and Physical Sciences Research Council (Grant EP/R005192/1).

\bibliography{article}

\begin{thebibliography}{10}
\expandafter\ifx\csname url\endcsname\relax\def\url#1{\texttt{#1}}\fi

\bibitem{pope00}
\Name{Pope S., Eccles P. \and Press C.~U.} \Book{Turbulent Flows} (Cambridge
  University Press) 2000.

\bibitem{richardson22}
\Name{Richardson L.} \Book{Weather Prediction by Numerical Process} (Cambridge
  University Press) 1922.

\bibitem{zakharov12}
\Name{Zakharov V., L'vov V. \and Falkovich G.} \Book{Kolmogorov Spectra of
  Turbulence I: Wave Turbulence} Springer Series in Nonlinear Dynamics
  (Springer Berlin Heidelberg) 2012.

\bibitem{monin54}
\Name{Monin A. \and Obukhov A.} \REVIEW{Contrib. Geophys. Inst. Acad. Sci.
  USSR}{151}{1954}{e187}.

\bibitem{foken06}
\Name{Foken T.} \REVIEW{Boundary-Layer Meteorology}{119}{2006}{431}.

\bibitem{barenghi14}
\Name{Barenghi C.~F., Skrbek L. \and Sreenivasan K.~R.} \REVIEW{Proceedings of
  the National Academy of Sciences}{111}{2014}{4647}.

\bibitem{kobayashi05}
\Name{Kobayashi M. \and Tsubota M.} \REVIEW{Phys. Rev.
  Lett.}{94}{2005}{065302}.

\bibitem{baggaley12}
\Name{Baggaley A.~W., Laurie J. \and Barenghi C.~F.} \REVIEW{Phys. Rev.
  Lett.}{109}{2012}{205304}.

\bibitem{thompson13}
\Name{Thompson K.~J., Bagnato G.~G., Telles G.~D., Caracanhas M.~A., dos Santos
  F. E.~A. \and Bagnato V.~S.} \REVIEW{Laser Physics
  Letters}{11}{2013}{015501}.

\bibitem{navon16}
\Name{Navon N., Gaunt A.~L., Smith R.~P. \and Hadzibabic Z.}
  \REVIEW{Nature}{539}{2016}{72}.

\bibitem{vivanco17}
\Name{Vivanco F. A.~J.} \Book{Investigations on momentum distributions and
  disorder in strongly out-of-equilibrium trapped {B}ose gases} , PhD Thesis,
  DOI: 10.11606/T.76.2017.tde-14092017-101126 (2017).

\bibitem{tsatos16}
\Name{Tsatsos M.~C., Tavares P.~E., Cidrim A., Fritsch A.~R., Caracanhas M.~A.,
  dos Santos F. E.~A., Barenghi C.~F. \and Bagnato V.~S.} \REVIEW{Physics
  Reports}{622}{2016}{1 }.

\bibitem{madeira20}
\Name{Madeira L., Caracanhas M., dos Santos F. \and Bagnato V.} \REVIEW{Annu.
  Rev. Condens. Matter Phys.}{11}{2020}{37}.

\bibitem{baggaley14}
\Name{Baggaley A.~W., Barenghi C.~F. \and Sergeev Y.~A.} \REVIEW{Phys. Rev.
  E}{89}{2014}{013002}.

\bibitem{navon19}
\Name{Navon N., Eigen C., Zhang J., Lopes R., Gaunt A.~L., Fujimoto K., Tsubota
  M., Smith R.~P. \and Hadzibabic Z.} \REVIEW{Science}{366}{2019}{382}.

\bibitem{henn09}
\Name{Henn E. A.~L., Seman J.~A., Roati G., Magalh\~aes K. M.~F. \and Bagnato
  V.~S.} \REVIEW{Phys. Rev. Lett.}{103}{2009}{045301}.

\bibitem{henn09_2}
\Name{Henn E. A.~L., Seman J.~A., Ramos E. R.~F., Caracanhas M., Castilho P.,
  Ol\'{\i}mpio E.~P., Roati G., Magalh\~aes D.~V., Magalh\~aes K. M.~F. \and
  Bagnato V.~S.} \REVIEW{Phys. Rev. A}{79}{2009}{043618}.

\bibitem{henn10}
\Name{Henn E., Seman J., Roati G., Magalhaes K. \and Bagnato V.}
  \REVIEW{Journal of Low Temperature Physics}{158}{2010}{435}.

\bibitem{seman11}
\Name{Seman J., Henn E., Shiozaki R., Roati G., Poveda-Cuevas F.,
  Magalh{\~{a}}es K., Yukalov V., Tsubota M., Kobayashi M., Kasamatsu K. \and
  Bagnato V.} \REVIEW{Laser Physics Letters}{8}{2011}{691}.

\bibitem{shiozaki11}
\Name{Shiozaki R., Telles G., Yukalov V. \and Bagnato V.} \REVIEW{Laser Physics
  Letters}{8}{2011}{393}.

\bibitem{vinen01}
\Name{Vinen W.~F.} \REVIEW{Phys. Rev. B}{64}{2001}{134520}.

\bibitem{barenghi14_2}
\Name{Barenghi C.~F., L{\textquoteright}vov V.~S. \and Roche P.-E.}
  \REVIEW{Proceedings of the National Academy of Sciences}{111}{2014}{4683}.

\bibitem{caracanhas13}
\Name{Caracanhas M., Fetter A.~L., Baym G., Muniz S.~R. \and Bagnato V.~S.}
  \REVIEW{Journal of Low Temperature Physics}{170}{2013}{133}.

\bibitem{dalfovo99}
\Name{Dalfovo F., Giorgini S., Pitaevskii L.~P. \and Stringari S.} \REVIEW{Rev.
  Mod. Phys.}{71}{1999}{463}.

\bibitem{abramowitz70}
\Name{Abramowitz M. \and Stegun I.} \Book{Handbook of Mathematical Functions
  with Formulas, Graphs, and Mathematical Tables} Applied mathematics series
  (U.S. Government Printing Office) 1970.

\bibitem{bewley06}
\Name{Bewley G.~P., Lathrop D.~P. \and Sreenivasan K.~R.}
  \REVIEW{Nature}{441}{2006}{588}.

\end{thebibliography}
\bibliographystyle{eplbib}

%
%
%
%

\end{document}